\documentclass[conference,a4paper]{APSIPA2023}
\usepackage{multirow}
\usepackage{graphicx}
\usepackage{amsmath}
\usepackage[psamsfonts]{amssymb}
\usepackage{amsxtra}
\usepackage{threeparttable}

\usepackage{geometry}
\usepackage{fancyhdr}

\fancypagestyle{firststyle} {
    \fancyhf{}
    \fancyhead[C]{2023 Asia Pacific Signal and Information Processing Association Annual Summit and Conference (APSIPA ASC)}

}

\begin{document}

\title{Exploring Isolated Musical Notes as Pre-training Data for Predominant Instrument Recognition in Polyphonic Music}

\author{%
\authorblockN{%
Lifan Zhong\authorrefmark{1}, Erica Cooper\authorrefmark{2}, 
Junichi Yamagishi\authorrefmark{2}, and Nobuaki Minematsu\authorrefmark{1}
}
\authorblockA{%
\authorrefmark{1}
Graduate School of Engineering, 
The University of Tokyo, Japan  \\
E-mail: \{zhong\_lifan, mine\}@gavo.t.u-tokyo.ac.jp  Tel/Fax:  +81-3-5841-6662}
\authorblockA{%
\authorrefmark{2}
National Institute of Informatics, Japan \\
E-mail: \{ecooper, jyamagis\}@nii.ac.jp Tel: +81-3-4212-2576}
}

\maketitle
\thispagestyle{firststyle}
\pagestyle{fancy}

\begin{abstract}
  With the growing amount of musical data available, automatic instrument recognition, one of the essential problems in Music Information Retrieval (MIR), is drawing more and more attention. While automatic recognition of single instruments has been well-studied, it remains challenging for polyphonic, multi-instrument musical recordings. This work presents our efforts toward building a robust end-to-end instrument recognition system for polyphonic multi-instrument music. We train our model using a pre-training and fine-tuning approach: we use a large amount of monophonic musical data for pre-training and subsequently fine-tune the model for the polyphonic ensemble. In pre-training, we apply data augmentation techniques to alleviate the domain gap between monophonic musical data and real-world music. We evaluate our method on the IRMAS testing data, a polyphonic musical dataset comprising professionally-produced commercial music recordings. Experimental results show that our best model achieves a micro F1-score of 0.674 and an LRAP of 0.814, meaning 10.9\% and 8.9\% relative improvement compared with the previous state-of-the-art end-to-end approach. Also, we are able to build a lightweight model, achieving competitive performance with only 519K trainable parameters.
\end{abstract}

\section{Introduction}

Music is an essential part of our lives. Nowadays, online music streaming services such as Apple Music and Spotify enable us to access an abundance of music recordings much more easily than before. 
At the same time, the massive amount of data brings new challenges:  how can we find the music we want efficiently? Music Information Retrieval (MIR) has emerged as a research field that focuses on extracting and inferring meaningful features from music, indexing music using these features, and developing various search and retrieval methods to make the world’s vast store of music easily accessible by individuals \cite{survey}.

As one of the important problems in MIR, instrument recognition shows great potential to contribute to real-world music applications, such as music recommendation and source separation \cite{hung2020multitask}. Though the past 30 years have seen great progress in the automatic classification of monophonic phrases or isolated notes, the task of instrument classification for polyphonic multi-instrument music, where the sound mixture becomes more complex, remains rather difficult. In this work, we tackle predominant instrument recognition, which focuses on identifying the main instruments in a given polyphonic music segment.\footnote{In this paper, ``polyphonic" means both polyphonic and multi-timbral, following the literature in this field.}

Generally speaking, training a robust statistical model for classification requires a large amount of high-quality annotated data. However, in the field of predominant instrument recognition, a lack of such annotated polyphonic musical data has been a constraint for training a robust classifier. First, domain knowledge is necessary for annotation in polyphonic musical mixtures, especially for complex ones. Second, well-produced music recordings help generalization but have copyright issues. On the contrary, 
monophonic sounds and isolated notes require relatively less effort to collect and label. But in terms of timbral characteristics, these are far from the music we listen to daily, which is consecutive, melodic, polyphonic, and not as ``clean'' as the isolated samples. Therefore, it would be interesting to discuss how monophonic musical data can help instrument recognition in a polyphonic setting. 

Some studies have touched on this. Bosch \MakeLowercase{\textit{et al.}} \cite{bosch2012comparison} made an early attempt by using source separation as a pre-processing step to divide polyphonic music signals into stems, and classified each stem through a separate model. Kratimenos \MakeLowercase{\textit{et al.}} \cite{kratimenos2021augmentation} experimented with four random mixing augmentation methods using the IRMAS training set \cite{bosch2012comparison}, which contains monophonic samples and polyphonic samples but all labeled with one predominant instrument, to improve the polyphonic instrument recognition system. 

In this paper, we look into this topic from another perspective. We investigate the effectiveness of using isolated musical notes as pre-training data for improving the predominant instrument recognition system. We adapt the model proposed in \cite{shi2022use} to classify instrument classes from raw musical waveforms and experiment with various data augmentation techniques to bridge the domain gap between isolated notes and polyphonic music. Furthermore, we investigate a weight-sharing method \cite{BOULCH201853} for reducing the parameters of our instrument encoder.

\begin{figure*}[t]
\begin{center}
\includegraphics[width=0.75\textwidth]{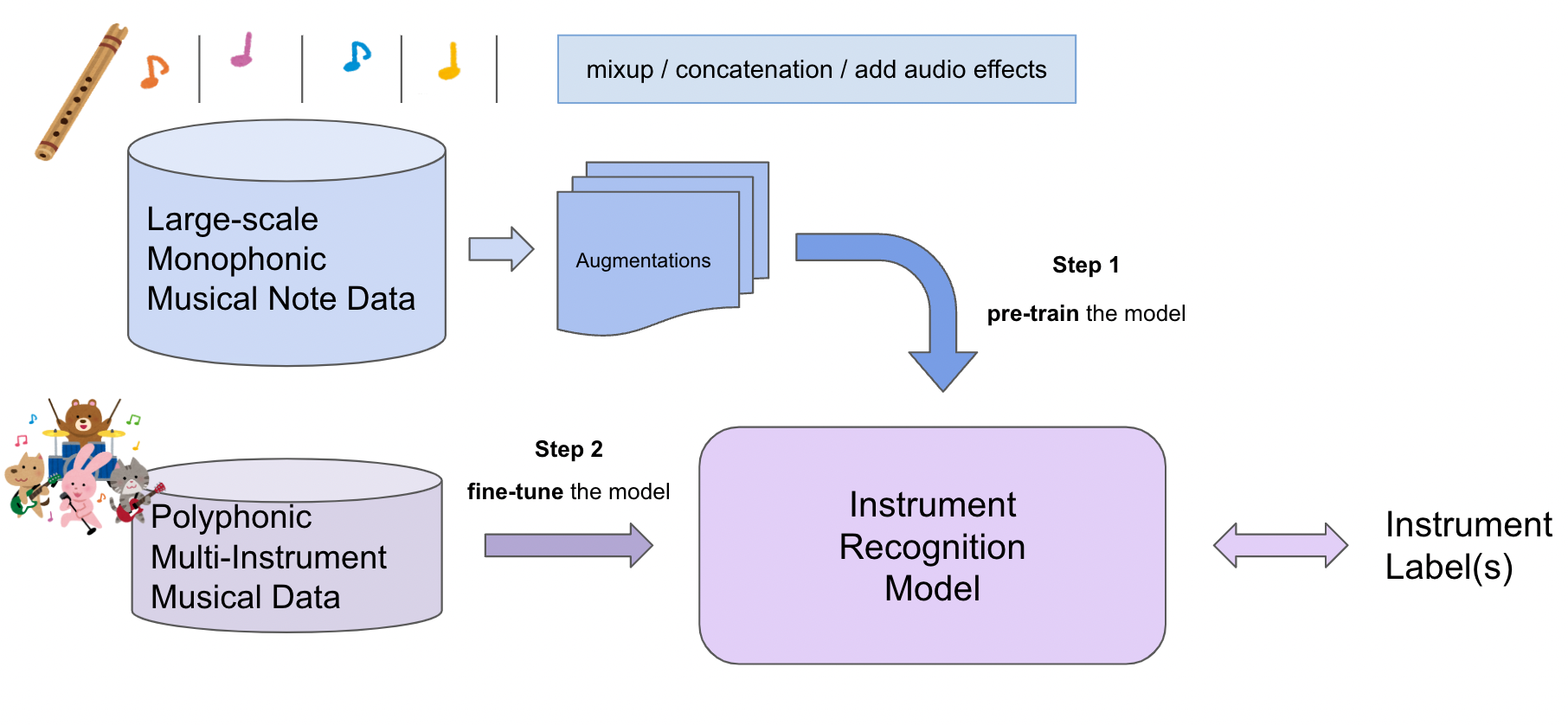}
\vspace{-5mm}
\end{center}
\caption{Overview of the proposed method.}
\label{fig:overview}
\end{figure*}

The remainder of the paper is structured as follows: Section \ref{related_work} gives a non-exhaustive review of the relevant research in the field of instrument recognition. Section \ref{method} describes the model architecture and data augmentation pipeline. Section \ref{exp} introduces the details of our experiment settings. Section \ref{results} presents our experimental results. Finally, Section \ref{conclusion} summarizes this work.

\section{Related Work}
\label{related_work}

Automatic classification of musical instruments has been studied since the mid-1990s. In 1995, Kaminsky \& Materka \cite{kaminsky1995automatic} trained instrument classifiers for four instruments in isolated notes. Eggink \& Brown \cite{eggink2003missing}\cite{eggink2004instrument} conducted two consecutive studies on instrument classification in polyphonic mixtures with models trained on monophonic audio. Livshin \& Rodet \cite{livshin2004musical} developed a real-time solo instrument classifier and also applied it directly to performance duets, demonstrating that the solo instrument classifier has the potential for multi-instrument classification.

As the automatic classification of isolated musical notes has been considered a solved problem \cite{lostanlen2018extended}, and monophonic music is not very applicable to real-world scenarios, more and more attention has shifted to instrument recognition in a polyphonic context. 
Fuhrmann \MakeLowercase{\textit{et al.}} \cite{fuhrmann2009scalability}\cite{fuhrmann2012automatic} took the first step towards the problem of predominant instrument recognition in polyphonic music by collecting a database and using hand-crafted features and support vector machines (SVM) to classify the music segments. Bosch \MakeLowercase{\textit{et al.}} \cite{bosch2012comparison} extended this research by using source separation as a pre-processing step. The dataset used in their research, the IRMAS (Instrument Recognition in Musical Audio Signals) dataset, later became the benchmark for predominant instrument recognition.

Recently, deep learning techniques have fostered more and more advances in polyphonic instrument recognition. In the task of predominant instrument recognition, Han \MakeLowercase{\textit{et al.}} \cite{han2016deep} continued the research on the IRMAS dataset by using VGG-styled convolutional neural networks (CNNs), and their system outperformed the previous SVM methods, marking a pioneering work in deep models for predominant instrument recognition. Pons \MakeLowercase{\textit{et al.}} \cite{pons2017timbre} made an attempt to modify the neural network structures based on domain knowledge, achieving similar scores to previous work with much fewer model parameters. Yu \MakeLowercase{\textit{et al.}} \cite{yu2020predominant} extended the CNN-based predominant instrument recognizer by combining additional spectral features and a multi-task approach based on instrument groups. Reghunath and Rajan \cite{reghunath2022transformer} applied Swin-Transformers to this task, and their input incorporates images of Mel-spectrograms as well as phase and tempo features. With an ensemble voting technique, a micro F1-score of 0.66 was achieved. Kratimenos \MakeLowercase{\textit{et al.}} \cite{kratimenos2021augmentation} proposed to randomly mix monophonic audio clips as augmented data samples to train neural networks for polyphonic instrument recognition. Avramidis \MakeLowercase{\textit{et al.}} \cite{avramidis2021deep} compared different architectures in an end-to-end fashion, using raw waveforms as model input. Zhong \MakeLowercase{\textit{et al.}} \cite{zhong2023transfer} investigated a cross-modal transfer learning approach for predominant instrument recognition using an ImageNet pre-trained vanilla ResNet-50 model.

\section{Methodology}
\label{method}

\begin{figure}[t]
\begin{center}
\includegraphics[width=0.4\textwidth]{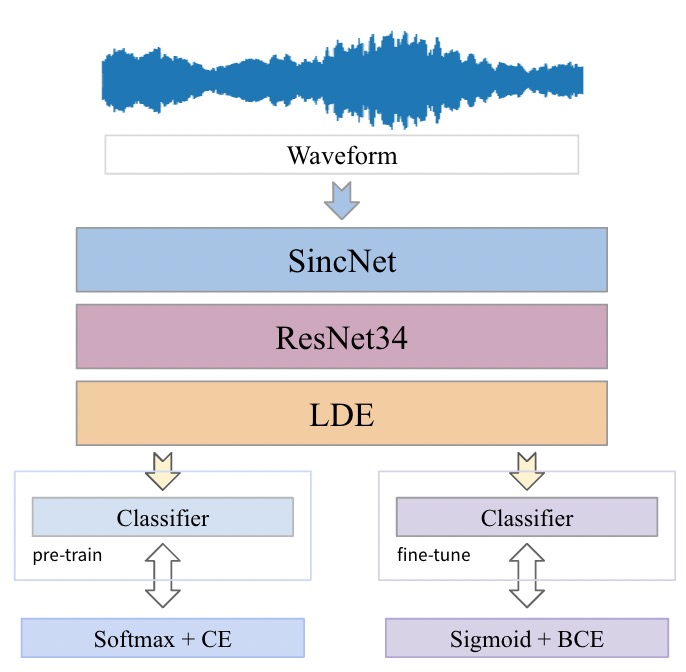}
\end{center}
\caption{Model Architecture.}
\label{fig:model}
\vspace*{-3pt}
\end{figure}

A high-level scheme of our proposed method is presented in Fig.\ \ref{fig:overview}. The training pipeline is as follows: First, we augment the monophonic musical note data by mixing, concatenating, and adding effects, to alleviate the domain gap between isolated notes and polyphonic recordings. The augmentations are done on-the-fly during the training process. Then we use the augmented data to pre-train the instrument recognition model. Finally, we fine-tune the pre-trained model for predominant instrument recognition using polyphonic, multi-instrument musical recordings. An overview of the instrument recognition model is shown in Fig.\ \ref{fig:model}. We will give a detailed explanation of the model and the augmentation methods in the following section.

\subsection{Model Architecture}
We adapt the instrument encoder proposed by Shi \MakeLowercase{\textit{et al.}} in \cite{shi2022use} as the main backbone. It consists of a learnable front-end, a CNN feature extractor, and a learnable pooling layer. The components of the backbone are described next. 

A SincNet \cite{ravanelli2018speaker} layer is used as the trainable front-end to extract the low-level features from the raw waveform. SincNet is implemented as a 1-D CNN layer. Instead of learning all the elements of the kernels, SincNet constrains its kernels to be a set of sinc functions in the time domain. Suppose $g$ is one SincNet kernel; then the operation is written as:
\begin{equation} 
g[n, f_1, f_2] = 2f_2 \, \rm{sinc}(2\pi f_2 n) - 2f_1 \, \rm{sinc}(2\pi f_1 n),
\end{equation} where $\rm{sinc}(x) = \sin(x) / x$. Then its Fourier transform is a rectangular bandpass filter in the frequency domain, which can be described as the difference between two low-pass filters:
\begin{equation} 
G[f, f_1, f_2] = \rm{rect}(\frac{f}{2f_2}) - \rm{rect}(\frac{f}{2f_1}),
\end{equation} where $\rm{rect}(\cdot)$ is the rectangular function. 
For such a parameterized filter, the only parameters to learn are $f_1$ and $f_2$, which are the low and high cutoff frequencies, respectively. 
Although in \cite{shi2022use}, a constant Q transform (CQT)-initialized SincNet provided better results, we did not find results showing improved polyphonic instrument recognition performance. Thus, the SincNet layer is initialized using the Mel-scale cutoff frequencies in our experiments, following the original SincNet paper.

We then use 34-layer residual convolutional networks (ResNets) \cite{he2016deep} to encode the low-level acoustic features to meta-level feature vectors. ResNets are CNNs which have identity mapping by shortcuts in every few stacked layers. 

Finally, we use a learnable dictionary encoding (LDE) layer \cite{cai2018novel} to aggregate CNN-encoded features as the instrument embeddings. In a nutshell, LDE learns a dictionary as a set of clustering centroids based on the weighted distance between every frame-level feature and every centroid.

\subsection{Data Augmentation}

We use the following three data augmentation strategies at the pre-training stage: 

\subsubsection{Concatenation} Some instruments cannot produce sustained notes. Thus, following \cite{shi2022use}, we randomly concatenate trimmed samples of the same instruments. By doing so, we also introduce note changes within an instrument in the generated samples, which is common in real-world recordings. 

\subsubsection{Mixup}Since multiple instruments may play at the same time in multi-timbral polyphonic music, in the pre-training phase, we adopt a mixup style training process \cite{tokozume2018learning} \cite{zhang2018mixup}, which trains the networks on convex combinations of pairs of examples and their labels. This process can be described as:
\begin{align} 
x_{mix} &= \lambda x_{i} + (1 - \lambda)x_{j}, \\
y_{mix} &= \lambda y_{i} + (1 - \lambda)y_{j},
\end{align} 
where $x_{i}$ and $x_{j}$ are two randomly selected audio waveforms of the same length, $y_{i}$ and $y_{j}$ are their labels respectively, $\lambda \in [0, 1] \sim \text{Beta} (\alpha, \alpha)$ is the random mixing ratio, $x_{mix}$ is the generated waveform, and $y_{mix}$ is the corresponding mixed label. In our implementation, we randomly select the audio samples regardless of the instrument class for mixing. Before combing the samples, we normalized them to the same Loudness Units relative to Full Scale (LUFS) level \cite{ITU-R}\footnote{https://github.com/csteinmetz1/pyloudnorm}. LUFS is a weighted loudness measure mimicking the relative sensitivity of the human ear to different frequencies in terms of perceived loudness.  

\subsubsection{Audio effects}As in reality, professionally-produced music often involves different kinds of audio effects such as reverb, delay, etc., we apply audio effect augmentation, including adding noise, delay, reverb, gain, pitch shift, and high-low pass for pre-training\footnote{https://github.com/Spijkervet/torchaudio-augmentations}. According to \cite{ramires2019data}, augmentation with audio effects commonly used in electronic music production improves the instrument recognition accuracy in both processed and unprocessed test sets.

\subsection{Fine-tuning for Polyphonic Multi-instrument Recordings}

Pre-training uses augmented monophonic note data, and its optimization criterion is to select only one most plausible instrument among multiple instruments. We therefore update the model with the softmax layer based on the cross-entropy loss with corresponding instrument labels.  

The criterion of the final task, on the other hand, is to select one or more principal instruments in a segment of the input polyphonic multi-instrument music recording, which is different from the pre-training criterion. Hence, for fine-tuning, we replace the softmax layer with a Sigmoid layer where each dimension corresponds to a probability for each instrument class and compute an element-wise binary cross-entropy loss so that more than one class may be predicted by setting a threshold value for the predicted probability of each instrument class.

\section{Experiments}
\label{exp}
\subsection{Data Preparation}
In our experiments, we pre-train our models on the NSynth dataset \cite{engel2017neural}, fine-tune the models with the IRMAS training data \cite{bosch2012comparison}, and perform the evaluation on the IRMAS testing data. Table \ref{table:summary} shows a summary of the two datasets.

\begin{table}[t]
    \caption[Brief Summary of NSynth dataset \cite{engel2017neural} and IRMAS dataset]{Summary of the NSynth dataset and the IRMAS dataset}
    \label{table:summary}
    \begin{center}
    \begin{tabular}{|l|ccc|}
        \hline
        Dataset & NSynth & IRMAS - train & IRMAS - test \\
        \hline
        \# Instruments & 1,006 & 11 & 11 \\
        \hline
        \# Samples & 305,979 & 6,705 & 2,874 \\
        \hline
        Duration per sample & 4 seconds & 3 seconds & 5 - 20 seconds \\
        \hline
        Total duration & 340.0 hours &  5.6 hours & 13.5 hours \\
        \hline
    \end{tabular}
    \end{center}
\end{table}

\subsubsection{NSynth}
The NSynth dataset is a collection of 305,979 audio snippets in 16 kHz, each lasting 4 seconds, containing an isolated monophonic musical note. Each note was held for the first 3 seconds, and for the last second, it was allowed to decay. Every note is different in instrument, pitch, or velocity and generated using one of 1,006 instruments from a commercial sample library. These 1,006 instrument labels are fine-grained, and each belongs to one of 11 instrument families: bass, brass, flute, guitar, keyboard, mallet, organ, reed, string, synth lead, and vocal. Since not all instruments play a full range of 88 pitches, NSynth contains an average of 65.4 pitches per instrument. 
However, in NSynth, some notes contain a pitch unusual for the particular instrument, thus producing nearly silent sounds that are hazardous to the classification. We therefore filter out these notes by applying an energy threshold before pre-training.
NSynth was released with an official split of train, validation, and test sets, and the instruments in the train set do not overlap with the ones in the validation set or test set. In our work, we formulate the pre-training process as a multi-class classification task in which the labels of samples in the validation set should be seen in the training phase. So we combine the train and validation sets in the official splits, resulting in a new train set with all 1,006 instruments. Then, 4\% of the new train set is separated as a new validation set. 

\subsubsection{IRMAS}
We use the IRMAS dataset to fine-tune and evaluate our instrument recognition model. This dataset consists of stereo recordings of professionally produced music excerpts with a sampling rate of 44.1 kHz, featuring varying instrumentation, playing styles, genres, audio quality, and production styles. Every sample in the dataset contains the segment-wise label of eleven pitched instruments, specifically cello (cel), clarinet (cla), flute (flu), acoustic guitar (gac), electric guitar (gel), organ (org), piano (pia), saxophone (sax), trumpet (tru), violin (vio), and human singing voice (voi). Note that the human voice was labeled as one of the instruments, while bass, drums, and percussion instruments were not annotated. Synthesizers, now essential ingredients in popular music, are not included in the annotations, either. IRMAS provides an official split of the dataset into training and testing sets. The training set comprises 6,705 3-second audio files, some are monophonic and others are polyphonic. But they all have only one predominant instrument label. The testing set includes 2,874 audio files, each lasting from five to twenty seconds and labeled with one or more predominant instruments. All the samples in the IRMAS dataset are downsampled to 16 kHz and converted to mono by taking the mean of the two stereo channels. Then, they are normalized to -12 LUFS.

We use the raw waveform signals as the input to our model. In the pre-training phase, only the first second of each sample in the augmented NSynth dataset is used as the fixed-length input for training and validation. There are two main reasons for doing so: first, as found in previous research \cite{han2016deep}, a 1-second duration is sufficient for the CNNs to capture the features of the instruments involved, and it was also found to be the optimal setting; second, in our preliminary experiments, we found that longer duration produced worse results. In the fine-tuning phase, for the IRMAS dataset, every 3-second audio sample in the training set is divided into three 1-second clips, resulting in 20,115 training samples. 15\% of the training samples are separated to form a validation set. However, the 3-second samples in the IRMAS training set are from 2,000 distinct songs. Obviously, within each song, the samples share similar timbral characteristics. So, in creating a validation set, we make sure that there are no overlaps with songs in the training split.

\subsection{Training Configuration}
In this study, all experiments are implemented with the PyTorch framework. The source code is available online\footnote{https://github.com/zhonglifan/Predominant-IR-NSynth-Pre-training}.

\subsubsection{Pre-training}The models are pre-trained using vanilla categorical cross-entropy loss on 2 GPUs with synchronized batch normalization \cite{Zhang_2018_CVPR}. We set the mini-batch size to 128 and trained the model for 30 epochs. We use Adam optimizer \cite{kingma2015adam}, and a weight decay of 0.0005.
The maximum value of the learning rate is set to 0.001. In training, the learning rate increases linearly from zero to the maximum value for the first three epochs and decreases to zero following a cosine curve.
We set label smoothing \cite{szegedy2016rethinking} to 0.05. Mixup is applied with a probability of 0.5 and $\alpha = 0.3$. Concatenation augmentation was applied with a probability of 0.5, and we sliced the first second of the generated samples as the input. The effect augmentation ratio was set to 0.3. We use the model weights of the last epoch as the starting point for fine-tuning.

\subsubsection{Fine-tuning} We fine-tune the models with the IRMAS training data. We use binary cross entropy loss and Sigmoid as the final activation. This is because our testing samples have one or more than one instrument labels. We set the mini-batch size to 64 and trained the model for 40 epochs on a single GPU.  We use Adam optimizer and a weight decay of 0.0005. The maximum value of the learning rate is set to 0.00025. To examine the effects of pre-training, we also train models from scratch for comparison. For the models trained from scratch, we set the learning rate to 0.0035, which is the optimal setting based on the results of the preliminary experiments. We believe this leads to a fairer comparison.
Again, the ``warm-up" training strategy is adopted: the learning rate increases linearly from zero for the first five epochs and decreases following a cosine curve. No data augmentations were implemented during fine-tuning since we didn't observe any improvements using augmentations. We use the model weights of the last epoch for evaluation.

\subsection{Evaluation Methodology}

\begin{figure*}[t]
\begin{center}
\includegraphics[width=0.75\textwidth]{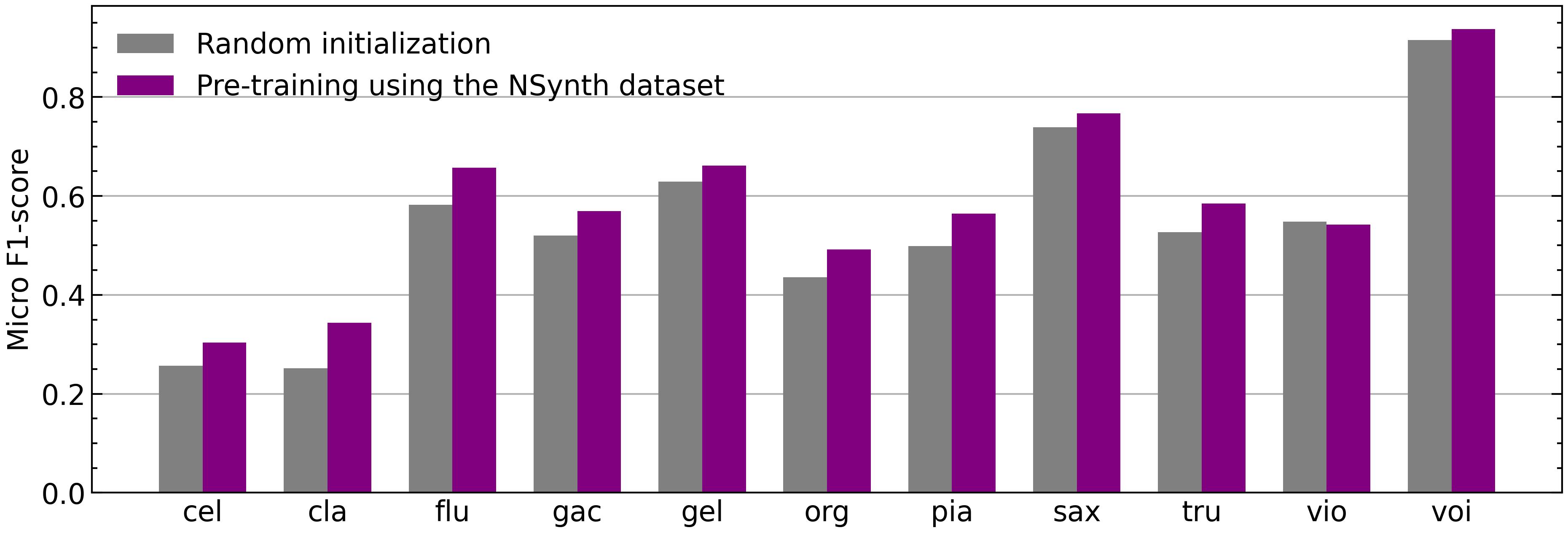}
\end{center}
\caption{Instrument-wise analysis: Training from random initialization \textit{vs.} with NSynth pre-training.}
\label{fig:instrument-wise-analysis}
\vspace*{-3pt}
\end{figure*}

We use the IRMAS testing set to evaluate our model, following the evaluation settings of previous works \cite{han2016deep}\cite{kratimenos2021augmentation}\cite{avramidis2021deep}. The testing samples in the IRMAS dataset are of varying lengths. To handle this, we divide each of them into 1-second fixed-length clips with 50\% overlap and take the average of clip-wise logits as the sample-wise predictions. 
As previously mentioned, there can be one or more predominant instrument labels, so generally, predominant instrument recognition is formulated as a multi-class multi-label classification task. We use the F1-score and Label Ranking Average Precision (LRAP) as evaluation metrics, consistent with previous research.

\subsubsection{F1-score}
F1-score is a commonly used metric in the field of instrument recognition. For each instrument label, F1-score is defined as the harmonic mean of precision and recall:
    \begin{equation}
    F1 = \frac{2 \cdot \text{Precision} \cdot \text{Recall}}{\text{Precision} + \text{Recall}},
    \end{equation}
and the corresponding precision and recall values for each instrument label are defined as:
    \begin{equation}
        \text{Precision} = \frac{\text{TP}}{\text{TP} + \text{FP}}, \quad 
        \text{Recall} = \frac{\text{TP}}{\text{TP} + \text{FN}},
    \end{equation} where TP is true positive, FP is false positive, and FN is false negative.

We calculate the averaged metrics over the 11 instrument labels to evaluate the overall model performance. To obtain the binary outputs, we compute the scores under a range of thresholds and report the optimal ones. The sample-wise predictions that exceed the threshold will be binarized as ones, i.e., the active labels. Since the label distribution of the IRMAS dataset is not balanced, both macro and micro-averaged metrics are calculated:

    \begin{equation}
        \text{P}_{macro} = \frac{1}{L}\sum_{l=1}^{L} \frac{\text{TP}_{l}}{\text{TP}_{l} + \text{FP}_{l}},
    \end{equation}

    \begin{equation}
         \text{R}_{macro} = \frac{1}{L}\sum_{l=1}^{L} \frac{\text{TP}_{l}}{\text{TP}_{l} + \text{FN}_{l}},
    \end{equation}

    \begin{equation}
        \text{P}_{micro} = \frac{\sum_{l=1}^{L}\text{TP}_{l}}{\sum_{l=1}^{L} (\text{TP}_{l} + \text{FP}_{l})},
    \end{equation} 

    \begin{equation}
        \text{R}_{micro} = \frac{\sum_{l=1}^{L}\text{TP}_{l}}{\sum_{l=1}^{L} (\text{TP}_{l} + \text{FN}_{l})},
    \end{equation}
where $l$ denotes the index of each label, and $L$ denotes the number of labels. respectively, The macro and micro averaged F1-scores are defined as:
    
    \begin{equation}
        F1_{macro} = \frac{2 \cdot \text{P}_{macro} \cdot \text{R}_{macro}}{\text{P}_{macro} + \text{R}_{macro}}, 
    \end{equation}

    \begin{equation}
        F1_{micro} = \frac{2 \cdot \text{P}_{micro} \cdot \text{R}_{micro}}{\text{P}_{micro} + \text{R}_{micro}}.
    \end{equation}
The macro average considers each class to have the same weight, whereas the micro average treats each instance as having equal weight, thus assigning more weights to the classes with more samples.

\subsubsection{LRAP}
Proposed in \cite{schapire2000boostexter}, LRAP is a metric for multi-label classification, and it was introduced to the instrument activity detection task by Gururani \MakeLowercase{\textit{et al.}} \cite{gururani2018instrument}. Suppose we have $N$ samples and $M$ labels. Given ground truth labels $y \in \left\{0, 1\right\}^{N \times M}$ and predictions $\hat{f} \in \mathbb{R}^{N \times M}$, it is defined as:

\begin{equation}
    LRAP(y, \hat{f}) = \frac{1}{N}
  \sum_{i=1}^{N} \frac{1}{||y_i||_0}
  \sum_{j:y_{ij} = 1} \frac{|\mathcal{L}_{ij}|}{\text{rank}_{ij}}, 
\end{equation} where $\mathcal{L}_{ij} = \left\{k: y_{ik} = 1, \hat{f}_{ik} \geq \hat{f}_{ij} \right\}$ and $\text{rank}_{ij} = \left|\left\{k: \hat{f}_{ik} \geq \hat{f}_{ij} \right\}\right|$. $|\cdot|$ computes number of elements of the set and $||\cdot||_0$ computes the number of nonzero elements in a vector. The intuition of this metric is that our model should assign a higher score to the true labels than the other ones. Thus, LRAP does not require a threshold to have binary outputs as F1-score does.

\section{Experimental Results}
\label{results}

We present the evaluation results on the IRMAS testing set. Note that we run each experiment three times with the same configuration except for the random seed, and report the mean value and standard deviation.

\subsubsection{Effects of NSynth Pre-training}
To examine the effects of NSynth pre-training, we train models from scratch with only the IRMAS training data. The comparison is shown in Table \ref{table:initialization}. From this table, we can see that NSynth pre-training strongly improves the performance of downstream task, i.e. predominant instrument recognition (from 0.634 to 0.674 for F1-micro). After analyzing the average performance over all the labels, we visualize the instrument-wise micro F1-score to get more insights into how the NSynth pre-training weights benefit predominant instrument recognition. As shown in Fig.\ \ref{fig:instrument-wise-analysis}, NSynth pre-training boosts the system’s performance in almost all the instruments except for the violin. 

\begin{table}[t]
\begin{center}
\caption{Training with random initialization \textit{vs.} with NSynth pre-training}
\label{table:initialization}
\begin{tabular}{|l|c|c|c|}
        \hline
        Initialization & F1-micro & F1-macro & LRAP\\
        \hline
        Random & 0.634 $\pm$ 0.0075	& 0.536	$\pm$ 0.0127 & 0.780 $\pm$ 0.0057 \\
        NSynth & 0.674 $\pm$ 0.0068 & 0.584 $\pm$ 0.0068 & 0.814 $\pm$ 0.0020\\
        \hline
\end{tabular}
\end{center}
\end{table}

\begin{figure}[t]
\begin{center}
\includegraphics[width=0.4\textwidth]{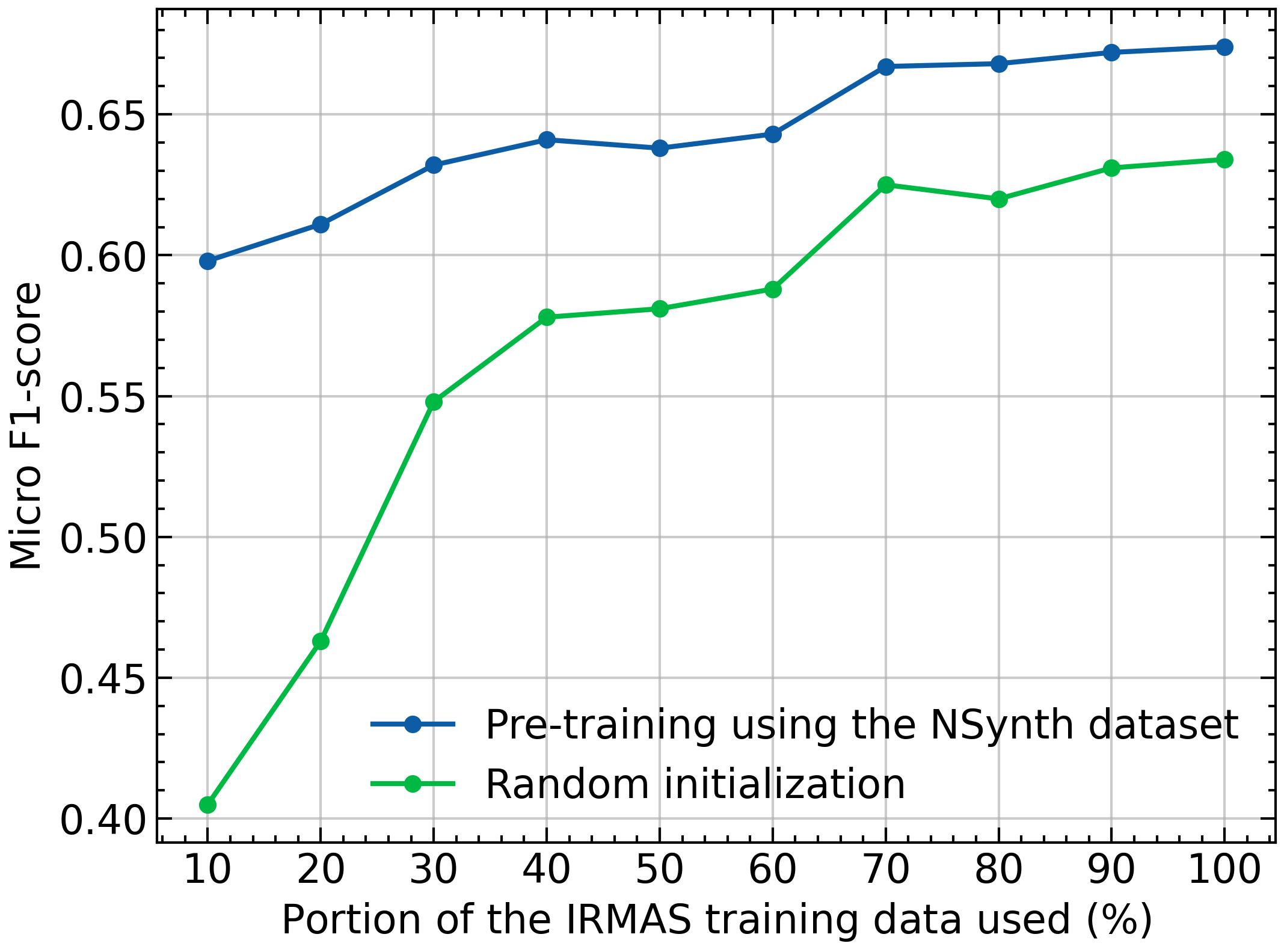}
\end{center}
\caption{Portion of the IRMAS training data used in the fine-tuning phase \textit{vs.} evaluation results on the IRMAS testing data.}
\label{fig:data-amount}
\vspace*{-3pt}
\end{figure}

NSynth itself is a large dataset, and its monophonic samples share similarities with polyphonic music data to a certain extent. Thus, we are interested to see how the effectiveness changes with the amount of the IRMAS training data used in the fine-tuning phase, in order to explore the effect of NSynth pre-trained weights a step further. Again, we report the results on the IRMAS testing set. Fig.\ \ref{fig:data-amount}  shows the micro F1-score obtained with respect to the amount of training data. Overall, the model benefits from NSynth pre-training regardless of the training data volume. The gap between pre-trained and randomly-initialized models is enlarged as the training data volume decreases. Surprisingly, with NSynth pre-trained weights and 10\% of the IRMAS training data, we can train a model that is reasonable for predominant instrument recognition.

\subsubsection{Effects of Data Augmentations}
Next, we ablate the data augmentation methods during the pre-training phase to examine their individual contributions to the results. The ablation results are shown in Table \ref{table:ablations}. While all augmentation techniques contribute to better performance, mixing two samples with soft labels has the most impact (by removing it, the micro F1-score drops from 0.674 to 0.657). And by pre-training without mixup and audio effects, the micro F1-score drops to 0.642.

\begin{table}[t]
\begin{center}
\begin{threeparttable}
\caption{Ablations of Pre-training Augmentation Methods}
\label{table:ablations}
\begin{tabular}{|l|c|c|c|}
        \hline
        Augmentations & F1-micro & F1-macro & LRAP\\
        \hline
        All  & 0.674 $\pm$ 0.0068 & 0.584 $\pm$ 0.0068 & 0.814 $\pm$ 0.0020 \\
        - mixup & 0.657 $\pm$ 0.0029 & 0.560 $\pm$ 0.0045 & 0.804 $\pm$ 0.0040\\
        - audio effect & 0.671 $\pm$ 0.0031	& 0.576 $\pm $0.0055 & 0.812 $\pm$ 0.0030\\
        - both\tnote{a}\quad & 0.642 $\pm$ 0.0050 & 0.535 $\pm$ 0.0031 & 0.791 $\pm$ 0.0037\\
        - concatenation & 0.670 $\pm$ 0.0012 & 0.576 $\pm$ 0.0015 & 0.813 $\pm$ 0.0013\\
        \hline
    \end{tabular}
\begin{tablenotes}
    \item[a] Without mixup and audio effects
\end{tablenotes}
\end{threeparttable}
\end{center}
\end{table}

\subsubsection{Single Predominant Instrument Identification}
\begin{figure}[t]
\begin{center}
    \includegraphics[width=0.482\textwidth]{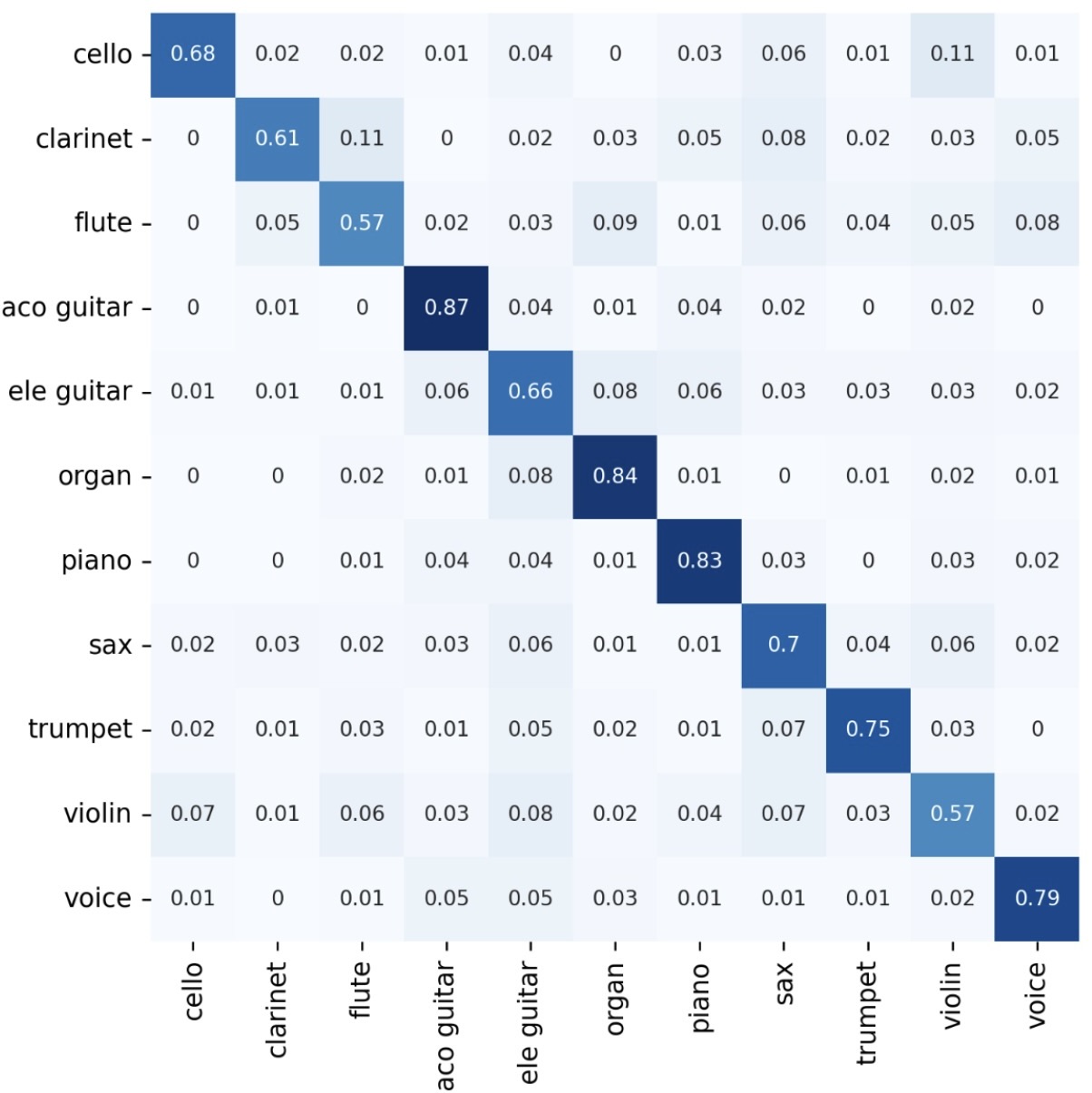}
    \caption{Confusion matrix of single predominant instrument identification. The columns are predictions and the rows are groun truth labels. Here, "aco guitar" and "ele guitar" stand for acoustic guitar and electric guitar, respectively.}
    \label{fig:cm}
    \vspace*{-3pt}
\end{center}
\end{figure}

\begin{figure}[t]
\begin{center}
    \includegraphics[width=0.482\textwidth]{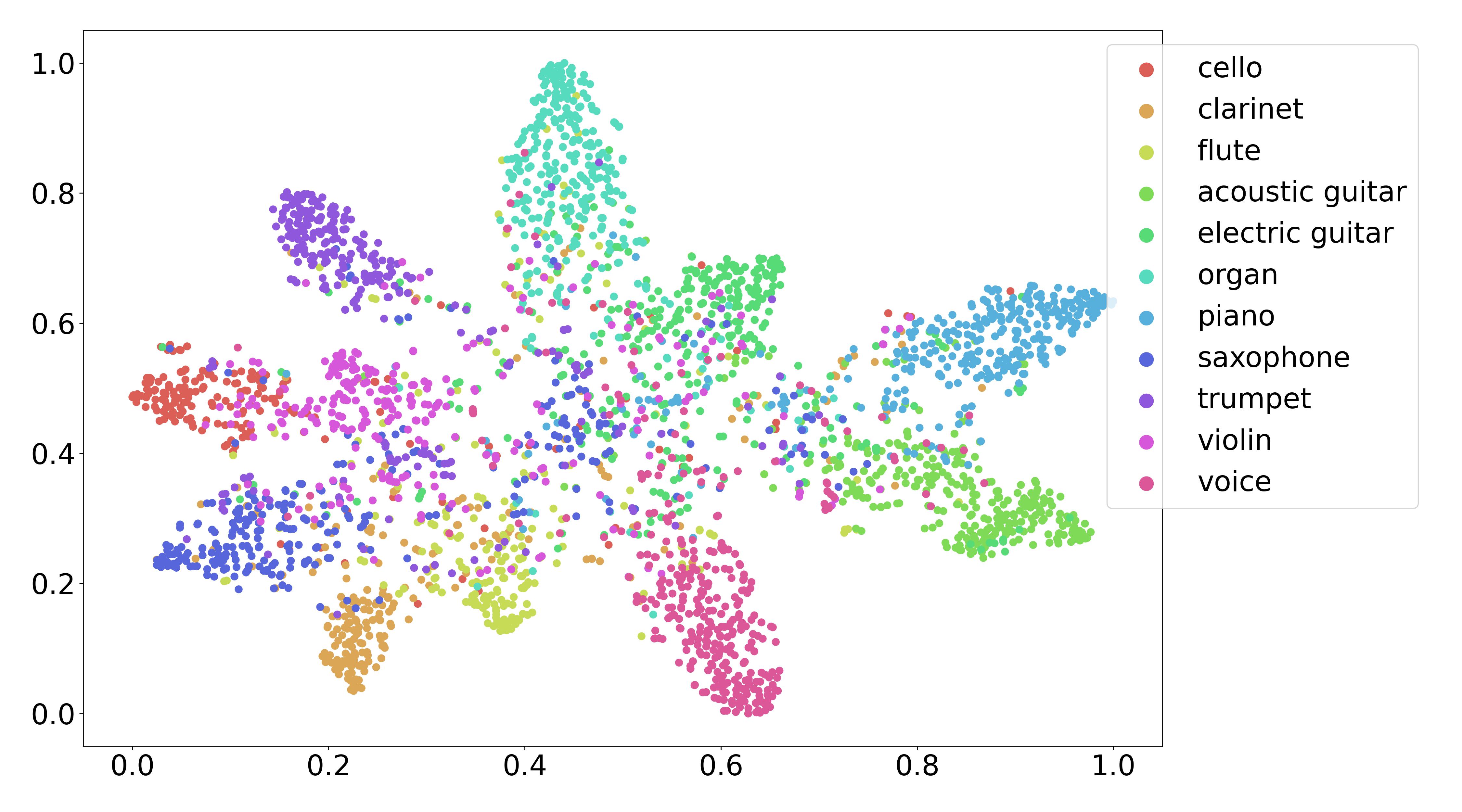}
    \caption{t-SNE visualization of the embeddings of the validation set.}
    \label{fig:tsne}
    \vspace*{-3pt}
\end{center}
\end{figure}

As in \cite{han2016deep}, we also analyze the model’s ability to identify the ``most" predominant instrument of the music segment. Here we report our results of the last epoch on the validation set. Note that we select the validation set to avoid similarity to the training split, so we believe that these results properly reflect the upper bound of our model’s generalization ability. The overall identification accuracy is 72.6\%. We present the confusion matrix in Fig.\ \ref{fig:cm}, and t-SNE visualization \cite{van2008visualizing} of the neural embeddings, explicitly the output of the penultimate layer, in Fig.\ \ref{fig:tsne}. From the results we can see that our model is prone to confuse the timbre of violin, flute, and clarinet with others.

\subsubsection{Parameter Reduction}

In the field of predominant instrument recognition, designing a reduced, lightweight model has been a research direction recently \cite{pons2017timbre} \cite{avramidis2021deep}. In our research, considering the potential applications of our system in DAW (Digital Audio Workstation) software, we want to reduce the parameters of the model, because when producing or mixing music on our local machines, we do not want the tagging system to take up too much of the limited hardware resources. 

\begin{figure}[t]
\begin{center}
    \includegraphics[width=0.482\textwidth]{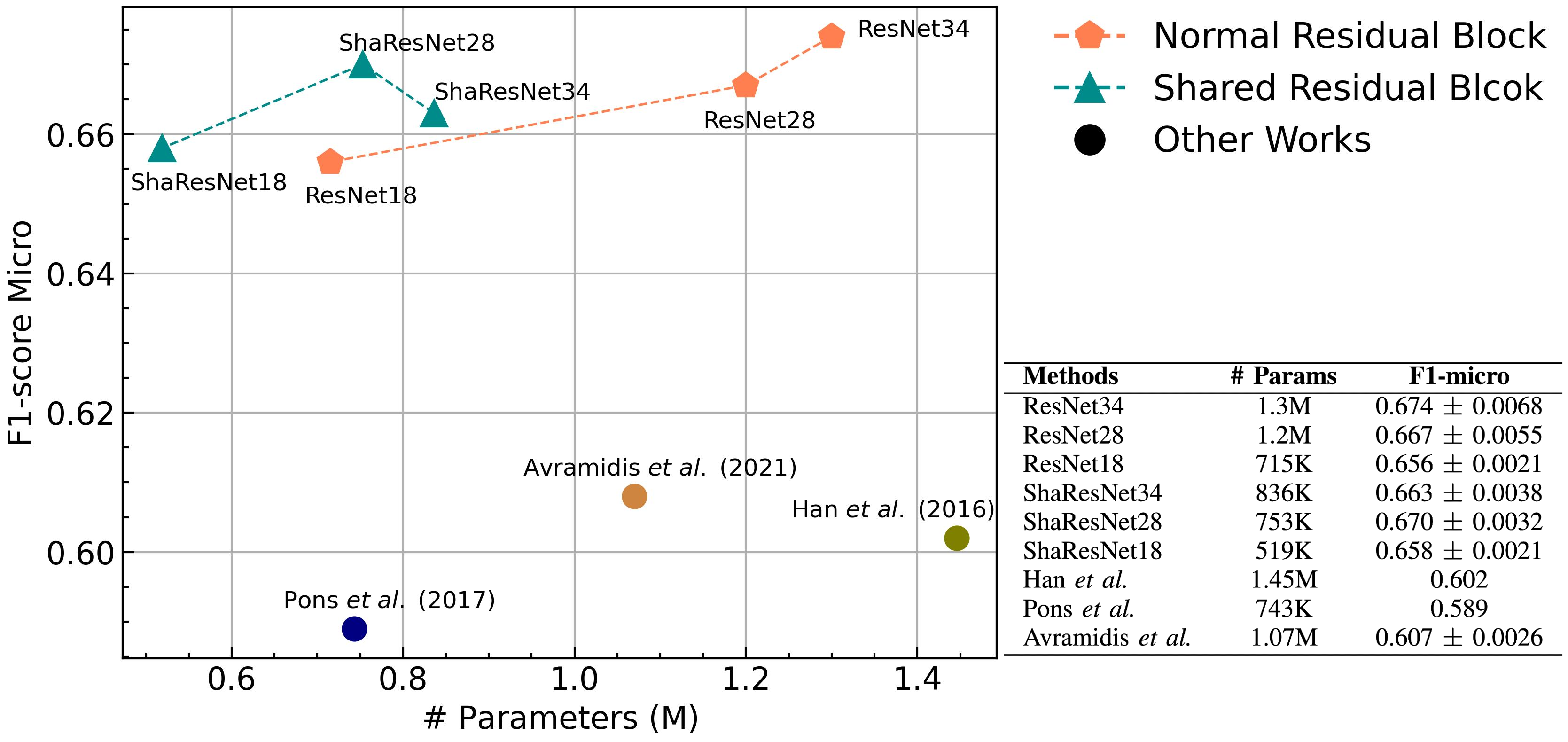}
    \caption{Comparison of evaluation performance between previous work and our reduced models.}
    \label{fig:parameters}
    \vspace*{-3pt}
\end{center}
\end{figure}

To reduce the parameters, we run experiments with backbones that are more shallow and further adopt a shared residual block strategy (ShaResNets) proposed in \cite{BOULCH201853} in consecutive residual blocks between two spatial downsampling layers. Fig.\ \ref{fig:parameters} presents the testing micro F1-score with respect to the models' number of parameters. We use $\left\{d_{1},d_{2},d_{3},d_{4}\right\}$ to denote the number of repetitive residual blocks in $n$-th section in ResNet. In this figure, ResNet34 is the model we start with, ResNet28 reduces the residual layers of backbone to $\left\{3,3,4,3\right\}$ and ResNet18's backbone has residual layers of $\left\{2,2,2,2\right\}$. Previous works are plotted as dots for comparison. The best performance is achieved by our base model, ResNet34, which is no surprise since it has the most parameters. Using shared blocks, we can reduce its number of parameters to 836K (64\%) with 0.011 (1.6\%) reduction in the F1 score. For ResNet28 and ResNet18, Sharing the weights achieves even higher performance than the original structures. ShaResNet28, which has 753K parameters, almost the same number of parameters as Pons's model in \cite{pons2017timbre}, has a competitive performance to our best model. Our smallest model, ShaResNet18, which only has 519K parameters, achieves a micro F1-score of 0.658, a surprising score considering its small size. Besides, from this figure, we can see a trend that as the number of parameters grows, the performance of normal residual networks grows, too. Thus we can infer that there is still room for improvement if we add more parameters to the model. 

\subsubsection{Comparison with Other Works}

\begin{table}[t]
\begin{center}
\begin{threeparttable}
\caption{Comparison of evaluation results on the IRMAS testing data}
\label{table:comparison}
\begin{tabular}{|l|l|c|c|c|}
        \hline
        Methods & Features & F1-micro & F1-macro & LRAP\\
        \hline
        This work  & Waveform & 0.674 & 0.584 & 0.814 \\
        Avramidis \MakeLowercase{\textit{et al.}} \cite{avramidis2021deep} & Waveform & 0.608 & 0.543 & 0.747 \\
        Kratimenos \MakeLowercase{\textit{et al.}} \cite{kratimenos2021augmentation} & CQT & 0.647 & 0.546 & 0.805 \\
        Zhong \MakeLowercase{\textit{et al.}} \cite{zhong2023transfer}\tnote{a} & Mel & 0.680 & 0.600 & 0.818 \\
        Reghunath \& Rajan \cite{reghunath2022transformer} & Mel\tnote{b} & 0.66 & 0.62 & - \\
        Yu \MakeLowercase{\textit{et al.}} \cite{yu2020predominant} & Mel & 0.661 & 0.569 & - \\
        Pons \MakeLowercase{\textit{et al.}} \cite{pons2017timbre} & Mel & 0.589 & 0.516 & -  \\
        Han \MakeLowercase{\textit{et al.}} \cite{han2016deep}\tnote{c} & Mel & 0.619 & 0.513 & -  \\
        \hline
\end{tabular}
\begin{tablenotes}
    \item[a] Here we show the results of single model. For the model ensemble in this work, the results were 0.688, 0.606, and 0.826 for F1-micro, F1-macro and LRAP respectively.
    \item[b] They fused Mel-spectrogram with modgdgram and tempogram as features.
    \item[c] Here we show their results in publication, but in preprint, the results were 0.602 and 0.503, for F1-micro and F1-macro respectively.
\end{tablenotes}
\end{threeparttable}
\end{center}
\end{table}

Table \ref{table:comparison} shows the micro-averaged F1-score, macro-averaged F1-score, and LRAP achieved by our proposed method and previous approaches. For all three evaluation metrics, the performance of our system produced the second best results. Our method, as an end-to-end classification approach that takes raw music data as input, outperforms the previous end-to-end system by 0.066 in micro F1-score, a 10.9\% relative improvement, which is noteworthy taking into account the difficulties of this task. Also, for all three metrics, our method shows a better performance than most of the previous methods that take time-frequency representations as inputs. \cite{zhong2023transfer} obtained a micro F1-score of 0.680 with a vanilla ResNet-50 whose number of parameters is 25.5M. But note that in this work, the best model only has 1.3M parameters, which is no more than 5.1\% of the former. Overall, this result indicates great potential in such end-to-end instrument recognition systems with trainable frontends.

\section{Conclusion}
\label{conclusion}

In conclusion, our study demonstrates the effectiveness of an end-to-end predominant instrument recognition system for polyphonic multi-instrument music. The key contributions of our research can be summarized as follows:

\begin{itemize}
    \item A pre-training and fine-tuning approach utilizing monophonic musical note data proves effective in predominant instrument recognition.

    \item Data augmentation techniques during pre-training help bridge the gap between monophonic data and real-world polyphonic music, contributing to the robustness of our model.

    \item Our best model achieves a micro F1-score of 0.674 and an LRAP of 0.814, marking a significant improvement of 10.9\% and 8.9\% relative to the previous end-to-end approach.

    \item A lightweight model with only 519K trainable parameters can still deliver competitive performance, demonstrating the potential for efficient deployment in various applications.
\end{itemize}

For future work, we will continue this research focusing on both model architecture as well as the challenges inherent in the data. First, we will improve the model performance by increasing the number of parameters, since the largest models we tested have only 1.3M parameters. Second, we would like to experiment with synthesized music rendered from MIDI scores as pre-training data. Moreover, we are also interested to explore this polyphonic instrument recognition problem in a zero-shot way.

\section*{Acknowledgements}

We would like to thank Ms.\ Xuan Shi from the University of Southern California for her useful comments.  This study is supported by JST CREST Grant Number JPMJCR18A6 and by MEXT KAKENHI grant 21K11951.

\end{document}